# GUPPY: Pythonic Quantum-Classical Programming

MARK KOCH, ALAN LAWRENCE, KARTIK SINGHAL, SEYON SIVARAJAH, and ROSS DUNCAN, Quantinuum, United Kingdom

We present ongoing work on GUPPY, a domain-specific language embedded in Python that allows users to write high-level hybrid quantum programs with complex control flow in Pythonic syntax, aiming to run them on actual quantum hardware.

```python
@guppy
def rx(q: Qubit, a: float) -> Qubit:
  # Implement Rx via Rz rotation
  return h(rz(h(q), a))
```

(a) Basic one-qubit gate in GUPPY

```python
@guppy
def teleport(
  src: Qubit, tgt: Qubit
) -> Qubit:
  # Entangle qubits with ancilla
  tmp, tgt = cx(h(Qubit()), tgt)
  src, tmp = cx(src, tmp)
  # Apply classical corrections
  if measure(h(src)):
    tgt = z(tgt)
  if measure(tmp):
    tgt = x(tgt)
  return tgt
```

(b) Quantum teleportation protocol in GUPPY

```python
@guppy
def rus(q: Qubit, tries: int) -> Qubit:
  for _ in range(tries):
    # Prepare ancillary qubits
    a, b = h(Qubit()), h(Qubit())

    b, a = cx(b, tdg(a))
    if not measure(t(a)):
      # First part failed; try again
      discard(b)
      continue

    q, b = cx(z(t(q)), b)
    if measure(t(b)):
      # Success, we are done
      break

    # Otherwise, apply correction
    q = x(q)

  return q
```

(c) Repeat-until-success protocol in GUPPY

Fig. 1. Example GUPPY programs.

## 1 Introduction

Python is by far the most popular programming language in the quantum domain. According to a recent poll [27], it is used by over 90% of researchers and practitioners in the field. Users appreciate Python's friendly syntax, supply of scientific libraries, and vast ecosystem. However, most Python frameworks are limited by the fact that they describe quantum programs at the abstraction level of circuits, with only limited support for algorithms with high-level control flow. For example, repeat-until-success protocols [14] use classical bits produced in real-time by quantum measurements. Hardware support for this feed-forward data processing will further increase with next-generation quantum devices entering the market [6]. Thus, we anticipate a rising demand for languages supporting these features.

Authors' Contact Information: Mark Koch, mark.koch@quantinuum.com; Alan Lawrence; Kartik Singhal; Seyon Sivarajah; Ross Duncan, Quantinuum, Cambridge, United Kingdom.

To satisfy this need, we introduce Guppy: a domain-specific language embedded in Python that allows users to write high-level hybrid quantum programs with complex control flow in Pythonic syntax. While most Python-based frameworks trace the Python interpreter (see Sec. 4.1) to build the representation of a quantum program, Guppy code is parsed separately and statically compiled to a novel quantum intermediate representation called Hugr [11], which can express and optimise these real-time quantum-classical programs (see Sec. 3). This gives us the flexibility to add new syntactic constructs, custom type checking, and better error messages.

As usual for embedded languages, Guppy programs are defined and compiled *within* an outer host program, in our case, a Python script. To indicate that a function should be handled by the Guppy compiler instead of the Python interpreter, users add the `@guppy` decorator to the function signature. While Guppy compilation happens within the Python interpreter, the compiled program may run at some other time independent of the Python runtime.

Before looking at Guppy's features in detail, we give a taste of what programming in Guppy feels like by walking through some examples. Figure 1a uses Guppy's `Qubit` type to build a basic quantum operation. Qubits in Guppy are *linear*, i.e., they cannot be copied or discarded. This allows us to catch common programming errors at compile time (see Sec. 2.4). Figure 1b implements the quantum teleportation protocol in Guppy. The classical corrections necessary for teleportation are naturally expressed in Guppy via a standard Python `if` statement. Finally, Figure 1c shows a more involved example, implementing the $V_3$ gate using the repeat-until-success algorithm by Paetznick and Svore [14]. It features a more complicated control flow involving looping and jumps based on measurement results, which are all easily expressible in Guppy's Pythonic syntax.

## 2 Features

### 2.1 Pythonic Control Flow

Guppy programs can contain arbitrary control flow composed of Python's `if`, `for`, `while`, `break`, `continue`, and `return` statements. In particular, control flow may be conditioned on measurement results, enabling real-time quantum-classical computation as shown in Figures 1b and 1c. Convenience features on boolean expressions, such as short-circuit evaluation and chained comparisons, are also supported.

### 2.2 Strong Typing

While typing in Python is strictly optional, Guppy code must type check, and the compiler rejects ill-typed programs. As such, users must annotate the type signature when defining a Guppy function. Furthermore, used variables must be previously assigned in all control-flow paths and must have a unique static type:

```
if b:                                  var = 42 if b else None
    var = 42                           use(var)  # Could be `int` or `None`
use(var)  # Not definitely assigned
```

These code fragments would be valid in Python but result in Guppy compiler errors since the variable var is not assigned (left) or may have different types (right) when b is falsy. We impose this restriction to deal with Python's dynamic nature when statically compiling Guppy code.

### 2.3 Basic Types and Operations

Guppy supports the standard Python types `bool`, `int`, `float`, and `None`. Furthermore, all operations, most built-in methods, and the corresponding "dunder" methods (e.g., `__add__`, `__eq__`, `__bool__`, etc.) are available. Python's numeric tower is faithfully represented with implicit coercions for arithmetic operations between different numeric types. The main difference from Python's regular semantics is that Guppy integers are limited to 64 bits, unlike Python's unbounded integers, with over- and underflows resulting in runtime errors.

## 2.4 Quantum Operations

Guppy features the type `Qubit` that quantum operations act on. Values of type `Qubit` are treated linearly, i.e., they cannot be copied or implicitly discarded. In this model, quantum operations take qubits as input and return them as output. For example, the Hadamard operation has the signature `h: Qubit -> Qubit`. Linear types allow us to catch programming mistakes at compile time instead of running into costly runtime errors:

```
q = Qubit()                          q = Qubit()
return cx(q, q)  # Multiple uses     h(q)  # Return value not used
```

Here, the programmer violated the no-cloning theorem by using the same qubit twice (left) and accidentally dropped a qubit by discarding a return value (right).

## 2.5 Collections

Guppy supports linear versions of Python's built-in lists and tuples. For example, lists of qubits can be used to represent qubit registers. The interface to interact with these lists containing linear data differs slightly from the usual Python API. For example, the `get` function on a qubit list not only returns the qubit at the given index but also a new list in which this qubit has been removed. To make working with linear lists easier, we provide a custom `apply` method that applies a function on given indices in a list. For example, the expression `qs.apply(cx, (i, j))` applies a CX gate on indices `i` and `j`, and returns a new list. This method is unsafe since the case `i == j` would result in a runtime error. This allows the user to suspend the linearity guarantee inside the register and instead rely on their own index reasoning. For example, a CX ladder could be constructed as follows:

```
n, qs = len(qs)
for i in range(n-1):
  qs = qs.apply(cx, (i, i + 1))
```

Guppy also supports Pythonic list comprehensions. For example, applying a Hadamard to every qubit in a register can be achieved via `[h(q) for q in qs]`. Finally, classical lists and tuples without the linearity restriction are also available in Guppy. However, at the moment, we only support immutable versions of these classical data types.

## 2.6 Higher-Order Functions

Functions in Guppy are first-class values typed via Python's `Callable` type constructor. For example, the higher-order `apply` function from Sec. 2.5 takes a function as its first argument. Functions can be used to specify subroutines and called by other functions defined in the same module. Furthermore, Guppy allows nested function definitions with captured (non-linear) variables and (mutual) recursion.

## 2.7 Python Interoperability

Since Guppy is embedded into Python, we may allow users to inject Python values into the Guppy program at compile time. To this end, we add a special expression `py(...)`, whose arguments are compile-time evaluated by the outer Python interpreter. In particular, the expression can use Python features that Guppy does not support and call out to other Python libraries:

```python
# Construct networkx graph in Python
import networkx as nx
g = nx.erdos_renyi_graph(20, 0.25)

@guppy
def apply_graph(qs: list[Qubit]) -> list[Qubit]:
  # Access Python graph from Guppy
  for i, j in py(g.edges()):
    qs = qs.apply(zz, (i, j))
  return qs
```

As long as the Python expression evaluates to a Guppy-compatible type, it can be used like any other value. In particular, pytket [22] circuits are interpreted as Guppy functions of type `list[Qubit] -> list[Qubit]` allowing seamless interoperability with existing pytket code. If the expression evaluates to an incompatible Python type T, the value is treated as an opaque data blob of type Opaque[T] and may only be passed to external functions. Variables defined at Guppy level may never be used in a py expression since their values are not available at compile time:

```python
var = 42  # Guppy variable
x += py(var + 1)  # `var` may not be used inside `py`
```

## 3 Compilation

The Guppy compiler, written in Python, compiles Guppy source code to the Hierarchical Unified Graph Representation (Hugr) [11]. Hugr is a directed graph-based intermediate representation designed to express quantum-classical programs and to allow optimisation of those programs within and across those domains. Hugr is used by version 2 of the TKET compiler [22, 26] to optimise quantum subroutines, with subsequent lowering to quantum-specific targets such as QIR [16]. Classical programs can be lowered and executed via an MLIR + LLVM pipeline [12].

Hugr is *hierarchical* in the sense that nodes in the graph can contain child nodes that form a subgraph. The logic of programs is represented in Hugr as dataflow graphs, where nodes are pure function calls, and edges represent data dependency. The partial ordering enforced by data dependency constrains execution order, so data parallelism is inherent.

The graph formalism of Hugr also allows for easy specification of local optimising rewrites as sub-graph pattern search followed by local replacement. We have implemented an efficient algorithm for matching many such patterns at once [13].

Within dataflow graphs, control flow can be represented using hierarchical structured control-flow nodes, which enables the local rewrite approach to be used across control-flow regions with Conditional and TailLoop nodes represent branching and iteration, respectively. The Hugr graph shown in Figure 2 uses Conditional nodes to express the teleportation example in Figure 1b.

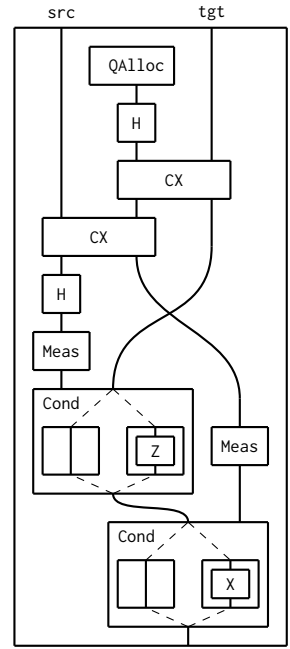

Fig. 2. Simplified Hugr representation of the teleportation example from Figure 1b

Dataflow graphs can also contain CFG nodes, the child nodes of which do not themselves obey dataflow but instead represent a Control Flow Graph (CFG). The nodes of the CFG are basic blocks, the child nodes of which (grandchildren of the CFG node) form dataflow graphs representing the logic for that basic block. CFG nodes can be used to capture arbitrary Guppy control flow directly and may be transformed into structured control flow at a later time [3].

Hugr also has native support for linear types, implemented by constraining node outputs in the graph to have exactly one dataflow edge connecting them to an input (a use of the value). A Guppy `Qubit` is compiled to such a linear type. Linear types allow local graph rewrites to be performed without violating the no-cloning theorem for qubits, such as two-qubit operations where both operands are the same qubit, as this would involve a double use of the linear type.

## 4 Discussion
### 4.1 Related Work

Various high-level quantum programming languages have been proposed in recent years. Q# [25] is a stand-alone language aimed at expressing hybrid quantum algorithms with complex control flow that go beyond mere circuit descriptions and compile to an expressive IR. It offers fewer safety guarantees, as qubits are represented as opaque pointers instead of Guppy's linear typing. However, there is a proposal to improve its safety [21].

Many quantum languages like Catalyst [10] and Qrisp [19], as well as frameworks like Cirq [7], ProjectQ [24], Pytket [22], PyQuil [23], and Qiskit [17] are based on Python. However, they all rely on tracing the Python interpreter to construct a program or circuit representation in the background and hence do not capture Python's native control-flow syntax. Thus, conditionals and looping must be expressed via higher-order combinators or other syntactic constructs. AutoQASM [5] improves on this using the `AutoGraph` module of `TensorFlow` [1]; however, it is more focused on providing an interface for low-level descriptions of quantum programs and compiles to QASM. The blqs framework [2] adapts the `AutoGraph` approach to instead generate Python quantum library objects such as Cirq and Qiskit. None of these languages and frameworks make use of linear types.

The idea of using linear typing to express quantum programs goes back to work by Selinger and Valiron [20] and has since been adopted in various languages like Proto-Quipper(s) [9, 18], QWire [15], and Qimaera [8]. However, all of these languages are based on the functional programming paradigm, which could be an entry barrier for programmers who prefer the imperative style of Python. Finally, Silq [4] is an imperative language featuring linear types and an intricate type system to support automatic uncomputation. However, it has a less accessible syntax and lacks the library ecosystem of Python.

### 4.2 Future Work

We added the unsafe `apply` method for linear lists so that users can carry out their own (potentially non-linear) index reasoning (see Sec. 2.5). We are investigating ways to offer this notion of local suspension of linearity as a built-in language feature via *classical references* to linear values. These non-linear references will cause runtime failure if the same reference is dereferenced multiple times. However, many open questions remain regarding the lifetime of references and finding a user-friendly syntax. Furthermore, Guppy currently only offers quantum gates as building blocks to construct programs. In the future, we plan to add more high-level features, e.g., automatically generating controlled and adjoint versions of quantum operations and facilities for automatic uncomputation of temporary qubits. We also want to support more of Python's built-in types, like strings, sets, and dictionaries. Finally, we want to allow users to define their own struct-like types (i.e., named tuples and data classes) and operations on them via Python's class syntax.